\documentclass[10pt,conference]{IEEEtran}
\IEEEoverridecommandlockouts
\usepackage{cite}
\usepackage{amsmath,amssymb,amsfonts}
\usepackage{algorithmic}
\usepackage{graphicx}
\usepackage{textcomp}
\usepackage{xcolor}
\def\BibTeX{{\rm B\kern-.05em{\sc i\kern-.025em b}\kern-.08em
    T\kern-.1667em\lower.7ex\hbox{E}\kern-.125emX}}
\begin{document}

\title{LLM-based policy generation for intent-based management of applications}
\makeatletter
\newcommand{\linebreakand}{%
  \end{@IEEEauthorhalign}
  \hfill\mbox{}\par
  \mbox{}\hfill\begin{@IEEEauthorhalign}
}
\makeatother

\author{\IEEEauthorblockN{Kristina Dzeparoska$^*$}
\IEEEauthorblockA{\textit{Department of Electrical and Computer Engineering} \\
\textit{University of Toronto}\\
Toronto, Ontario \\
kristina.dzeparoska@mail.utoronto.ca}
\and
\IEEEauthorblockN{Jieyu Lin$^*$}
\IEEEauthorblockA{\textit{Department of Electrical and Computer Engineering} \\
\textit{University of Toronto}\\
Toronto, Ontario \\
jieyu.lin@mail.utoronto.ca}
 \linebreakand
\IEEEauthorblockN{Ali Tizghadam}
\IEEEauthorblockA{\textit{Department of Electrical and Computer Engineering} \\
\textit{University of Toronto}\\
Toronto, Ontario \\
ali.tizghadam@utoronto.ca}
\and
\IEEEauthorblockN{Alberto Leon-Garcia}
\IEEEauthorblockA{\textit{Department of Electrical and Computer Engineering} \\
\textit{University of Toronto}\\
Toronto, Ontario \\
alberto.leongarcia@utoronto.ca}

}

\maketitle

\begin{abstract}
Automated management requires decomposing high-level user requests, such as intents, to an abstraction that the system can understand and execute. This is challenging because even a simple intent requires performing a number of ordered steps. And the task of identifying and adapting these steps (as conditions change) requires a decomposition approach that cannot be exactly pre-defined beforehand. To tackle these challenges and support automated intent decomposition and execution, we explore the few-shot capability of Large Language Models (LLMs). We propose a pipeline that progressively decomposes intents by generating the required actions using a policy-based abstraction. This allows us to automate the policy execution by creating a closed control loop for the intent deployment. To do so, we generate and map the policies to APIs and form application management loops that perform the necessary monitoring, analysis, planning and execution. We evaluate our proposal with a use-case to fulfill and assure an application service chain of virtual network functions. Using our approach, we can generalize and generate the necessary steps to realize intents, thereby enabling intent automation for application management.
\end{abstract}

\def\thefootnote{*}\footnotetext{equal contribution}\def\thefootnote{\arabic{footnote}}
\footnote{This article has been accepted for publication in 2023 19th International Conference on Network and Service Management (CNSM), 3rd International Workshop on Analytics for Service and Application Management (AnServApp 2023), DOI: 10.23919/CNSM59352.2023.10327837}
%\footnote{DOI: 10.23919/CNSM59352.2023.10327837}

\section{Introduction}\label{intro}
The growing heterogeneous and distributed resources that support the plethora of services and applications can be challenging to manage, in particular considering dynamic environments and strict application requirements such as availability, security, and reliability. Human-based operations are prone to error, and time- and cost-sensitive. Automation is highly desirable to relieve administrators from repetitive and tedious management tasks, thereby keeping OpEx low, and focusing their attention on more complex problems. Recent advances in ML and AI can be leveraged to simplify these management efforts.

An intent defines a set of operational goals (that a system should meet) and outcomes (that a system should deliver), without specifying how to achieve or implement them \cite{rfc9315}. Intent-based networking (IBN) or intent-based management seek to automate network and management tasks by allowing the system to accept and realize user intents. Intents bring two main requirements that are categorized by functionality into intent fulfillment and intent assurance. The former includes: functions such as translation, decomposition and execution; abstractions that capture system hierarchy (from top-level applications to low-level infrastructure devices) and provide a logical view to support the intent-related functions. To ensure that intents are continuously met during their life-cycle, and to prevent "intent drift" \cite{rfc9315}, intent assurance requires the system to monitor and adapt the intent deployment accordingly. Finally to close the loops for assurance and fulfillment, negative feedback control loops can couple the required steps for the necessary measurements, decisions and control. 

Large Language Models (LLMs) provide powerful capabilities for Natural Language Processing (NLP) tasks, such as understanding, generating, and classification of text data. This makes LLMs attractive as a solution in translation and decomposition in intent processing. Moreover, LLM models are trained on massive datasets, which allows a model to learn and understand the contextual relationship in natural language data. This is beneficial for intent-based management because is provides the ability to learn and adapt the steps required to realize intents. In this paper, we leverage the few-shot learning capability of LLMs to generate progressively the steps to realize an intent. This approach can generalize to unseen intents and support dynamic environments, where pre-defined sequences of actions would fail.

We use a policy-based approach to capture and model the relevant abstractions at each level of the management hierarchy. Our goal is to automatically decompose intents into a sequence of policies that when executed will deploy the intent. Our focus is on the capabilities of generic large language models for intent-based application management.

We developed "Emergence", an intent based management system. In order to realize the notion that an intent specifies "what the user wishes the system to do, without specifying how", we implemented an LLM pipeline for progressive decomposition of intents into policies, and then use mapping functions for policy-to-API resolution. Through this pipeline, Emergence determines the "how" part of the intent. This allows Emergence to then proceed with intent deployment, and to support intent fulfillment and assurance. 

We use a language model to understand natural language data and context in order to derive the steps to realize an intent. These steps enable our system to gather the necessary monitoring data, analyze the data based on current environment conditions, and accordingly create and execute a plan to deploy the intent onto the infrastructure. To link these steps and execute the policies, we use feedback control loops (MAPE-K, monitor-analyze-plan-execute and knowledge \cite{1160055}). 

We evaluate Emergence in an intent use-case that involves a virtual network function (VNF) service chain with high availability, consisting of a deep packet inspection tool and a load-balancer. Our use-case demonstrates both intent fulfillment and assurance, by taking appropriate actions when issues are detected to ensure that the intent is continuously met during its life-cycle. 
%We evaluate Emergence in an intent use-case that involves a virtual network function (VNF) service chain consisting of a deep packet inspection tool and a load-balancer, a health check service to monitor and notify if any issues are detected, and a self-healing ability. Our use-case demonstrates both intent fulfillment and assurance, by taking appropriate actions when issues are detected to ensure that the intent is continuously met during its life-cycle. 

Our contributions in this paper are:
\begin{itemize}
    \item Propose the use of LLMs with few-shot learning for progressive decomposition of intents into policies to support intent-based application management.
    \item Evaluate the system on a cloud testbed and demonstrate intent fulfillment and assurance.
    \item Discuss opportunities of generic LLMs for intent-based management.
\end{itemize}

\section{Background and Related Work}\label{background}

Research in intent-based systems has focused on different requirements, scopes, architectures, and approaches for the intent-related tasks. Efforts by standardization bodies such as Internet Engineering Task Force (IETF), European Telecommunications Standards Institute (ETSI), International Telecommunication Union (ITU), and TM Forum, as well as in academia have considered the use of ML/AI, closed control loops, abstractions (formal intent languages, policy models, etc.). Recent surveys discuss these research efforts, requirements, and challenges \cite{9925251, 9128422, 8968429}.

Natural-language processing (NLP) is required to translate and formalize intents. Recurrent neural networks, such as long short-term memory (LSTM), have been used to translate intents \cite{jacobs2021hey, 9687516, 9838633, yang2023smart, vedula2020open}. These models extract the necessary information from the intent, which is then mapped to an abstraction model or language that the system understands in order to process the intent. For example, LUMI \cite{jacobs2021hey} uses bidirectional LSTMs to extract entities and translate intents into an intent language (NILE), and then into configuration commands. LUMI has a chat-based interface (Google Dialogflow) for users to express their intents to the system. Other chat-based proposals include: iNDIRA \cite{kiran2018enabling} uses NLP to construct semantic RDF graphs to understand, interact, and create the required network services; EVIAN \cite{mahtout2020using}, an extension of iNDIRA, uses the RASA NLP tool to translate intents. RASA is also used in \cite{10175491}, and Dialogflow is used in \cite{9492554} and \cite{9838633}. 

Large language models (LLMs) such as GPT (\cite{openai2023gpt4, brown2020language}), PaLM \cite{chowdhery2022palm} and LLaMA \cite{touvron2023llama} are transformer-based models with impressive capabilities for NLP tasks. These models are trained using massive amounts of text data, and have billions of parameters. While both LSTMs and LLMs handle context in text, LLMs typically outperform LSTMs in large-scale language tasks due to a self-attention mechanism \cite{vaswani2017attention}, that provides a more flexible way of handling long-range dependencies in text. Moreover, LLMs are capable of few-shot learning \cite{brown2020language}, which allows the model to learn to perform a new task by simply training the model with several examples for a given task at hand. ChatGPT by OpenAI uses the Generative Pretrained Transformer (GPT) language model to generate responses to natural language inputs. Although these models are not trained for intent-based management tasks, they can be useful tools for intent processing due to their capabilities for few-shot learning. In \cite{lin2023appleseed}, we explored this capability to decompose intents into a set of Python APIs. Here, we leverage LLMs for intent to policy decomposition. 

Abstraction languages allow the modeling of an intent to a formal model that the system can understand. Languages proposed for intent modeling include, NILE \cite{jacobs2021hey}, SNIL \cite{9838633}, LAI \cite{tian2019safely}. We use the formal policy framework from our prior work,\cite{dzeparoska2021towards} that allows us to model policies at different levels of abstractions across the system hierarchy. To promote the adoption of our policy framework, we created a mapping to the Metro Ethernet Forum Policy Driven Orchestration Model (MEF PDO) \cite{MEF_PDO}, which is extensive and YANG-defined to support automation.

\section{Methodology}\label{method}

We now describe the Emergence system for management of intent-based applications. In this paper we focus on the LLM pipeline to decompose intents into a policy-based abstraction that can be mapped to APIs for intent execution and deployment. Figure \ref{pipeline} gives an overview of the pipeline and its three stages. Each uses an LLM with few-shot learning. The first stage classifies intents into types known by the system. The second stage decomposes the  intent and type into policies. The third stage validates the policies for omissions or errors in the policy format, as well as for the correct ordering of policies. Before discussing the pipeline in detail, we describe our policy-based abstraction.

\subsection{Policy model functional abstraction}
We use a policy-based approach to provide the abstractions needed at each level of the system hierarchy. We defined a formal policy framework in \cite{dzeparoska2021towards}, to model different types of policies (e.g., utility, goal, action) at different levels of abstractions, and detect and resolve conflicts across the hierarchy. 

\begin{figure*}[t]
   \centering
   \includegraphics[width=\textwidth]{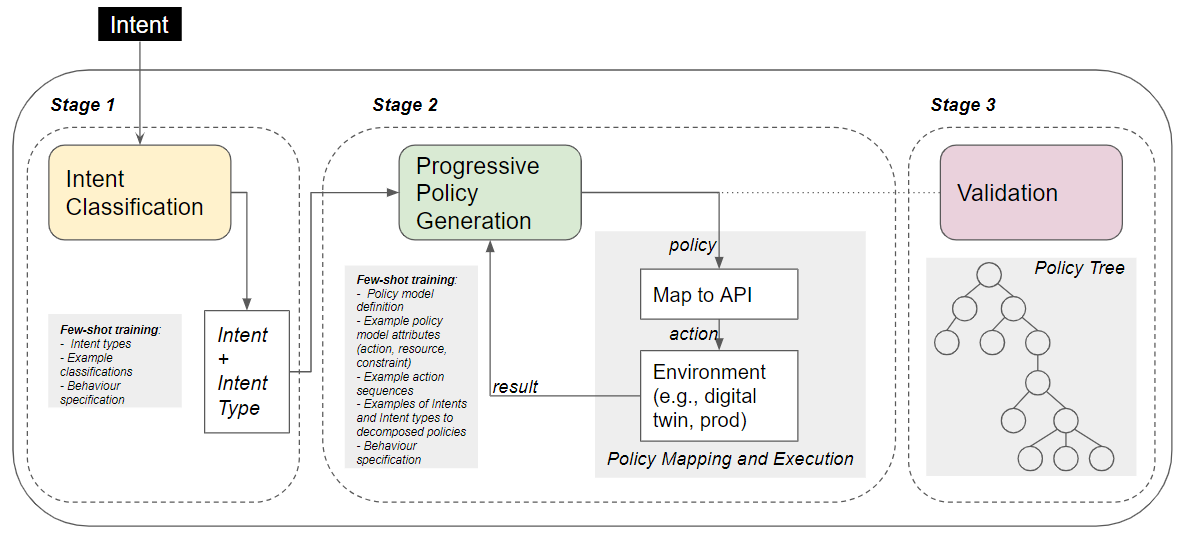}
   \caption{Pipeline overview: 1) classify intents to known intent types, 2) progressively decompose intents and generate policies: map each policy to an API, execute the API and return the result to the LLM, 3) validate the policies. The resulting policy tree represents the sequence of derived policies from the intent. % which will be executed using a closed control loop.
   }
   \label{pipeline} 
\end{figure*}

Our model formally defines Policy $\vec{P}$ as: %$\vec{P}=(D, E, A, \vec{C})$
\begin{equation}
\vec{P}=(D, E, A, \vec{C}).
\label{eq:policyModel}
\end{equation}
\noindent where $D$ denotes the policy definer that defines the policy, $E$ is an entity or a group of entities that enforce the policy, and $A$ is an action. $\vec{C}$ is a vector of constraints that apply to action $A$ with regards to resources ($\vec{R}$), temporal ($\vec{T}$) and spatial ($\vec{S}$) constraints, defined as: 
\begin{equation}
    \vec{C}=(\vec{R}, \vec{T}, \vec{S}).
    \label{eq:constraint}
\end{equation}
For each policy $\vec{P}$, we define policy metadata that includes the policy ID, domain, expiration date, priority, and autonomic permission. To support implementation of our policies in order to execute and deploy intents, we have created mapping functions to convert our policies to our APIs and to the MEF PDO model.

\subsection{Intent Decomposition using few-shot learning}\label{decomp}
Our goal is to decompose intents into policies with actions that correspond to a MAPE-K closed control loop. We use our policy model as an abstraction and train a generic LLM (OpenAI's ChatGPT) using few-shot learning to generate policies for user intents. In the few-shot part, we pass to the LLM an input message that generally contains: definitions (e.g., policy model) and descriptions, examples of intents and their corresponding intent types and policy sequences, and LLM behaviour specification (e.g., to act as a MAPE-K sequential policy generator). We find that LLMs can generalize well and learn to decompose intents, even with just a couple of intent to policy examples.

The generated policies represent the ordered sequence of policies to process and deploy an intent at a given time. We refer to this sequence as a policy tree (an example tree is shown in Figure \ref{pipeline}). As an example, for an intent that calls for the creation of a small virtual machine, the following policy (one of the required policies) is needed to check for availability of 1 small virtual machine in Domain1: $P = (avail, vm, zone=Domain1, size=small, count=1)$. To train the LLM we represent the policy as a JSON object as follows: $\{"action":"avail:, "resource":"vm", "zone":"Domain1", "size":"small", "count":1\}$. This allows us to train the LLM to output the policies in the same format (key-value pairs), making it easier to integrate them into code. For example, a dictionary data structure in Python allows us to map the policy action to an API and include the remaining policy attributes. We note that we purposefully omit the Definer and Enforcer attributes from the policies for clarity. The Definer is the user or application that creates or manages the intent, and the Enforcer is the MAPE component that enforces the policy. For example, for the above policy, the Definer is \textit{"Administrator"} and the Enforcer is \textit{"Analyze"}. 

\subsubsection{Intent Classification Stage}
The first pipeline stage uses an LLM to classify intents into one or more intent types supported by our system. To train the LLM with few-shot learning, we provided the intent types supported in our system, examples of intents and their corresponding intent types, and LLM behaviour specification (e.g., to act as an intent classifier). An intent may contain one or more intent types. For example, the intent: \textit{"Create a small monitored VM in domain 1."} is classified to the following types: \textit{create resource, schedule health check}. The first type is for the set of policies to create the VM. The second type is for the health check, since a VM cannot be monitored without having this check. Other intent types include: \textit{deploy service, start service, stop service, run service, discover resource, collect resource, publish resource, validate resource}, etc. Figure \ref{classify} shows the output of the classification results for three different intents.

\begin{figure}[h!]
   \centering
   \includegraphics[width=\columnwidth]{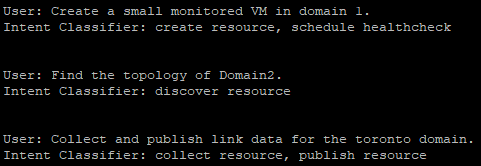} %newresults4-6.PNG
   \caption{Example intent to intent type classification for the first stage.}
   \label{classify} 
\end{figure}

\subsubsection{Progressive intent decomposition Stage}
The second pipeline uses an LLM to progressively decompose intents into policies based on each policy's execution result. The LLM is trained with few-shot learning to learn the steps (i.e., actions) required to deploy an intent, in a MAPE-based fashion. For this, we provide the LLM with the policy definition, system supported actions, and example resources and constraints. Examples are shown in Figure \ref{examples}. We also provide typical sequences of actions required for the different intent types our system supports. For example, for the intent type \textit{create resource}, the sequence of actions per MAPE component is: \textit{Monitor=[get], Analyze=[avail], Plan=[reserve], Execute=[create, validate]}. To help the model learn how to apply the above, we provide a couple of examples of intents, with their respective intent types and policies. In Figure \ref{training} we show a few-shot training example that is provided to the LLM to learn the intent to policy progressive decomposition process. Lastly, we specify the model's behaviour by instructing the model to act as a MAPE-K policy generator that outputs policies progressively, and uses the provided policy execution result to determine the next policy.

In our implementation, once the LLM generates a policy, we use mapping functions to convert the policy into an API. We then execute the API, and provide the result to the LLM. Next the LLM proceeds with the next policy until it derives the complete working sequence of actions (i.e., policy tree). Depending on the results and current conditions, the LLM can either conclude with an "END" message (indicating a successful intent completion), or an "ERROR" message. 

\subsubsection{Intent Validation Stage}
The last stage uses an LLM to validate the policy tree obtained from stage 2. We trained the validation LLM by modifying the prompt used for the decomposition LLM (stage 2), and included examples of incorrect policy sequences, and example corrections. This trains the LLM to look for any omissions or errors in the policy attributes, as well as incorrect policy sequences. 

\begin{figure}[h!]
   \centering
   \includegraphics[width=\columnwidth]{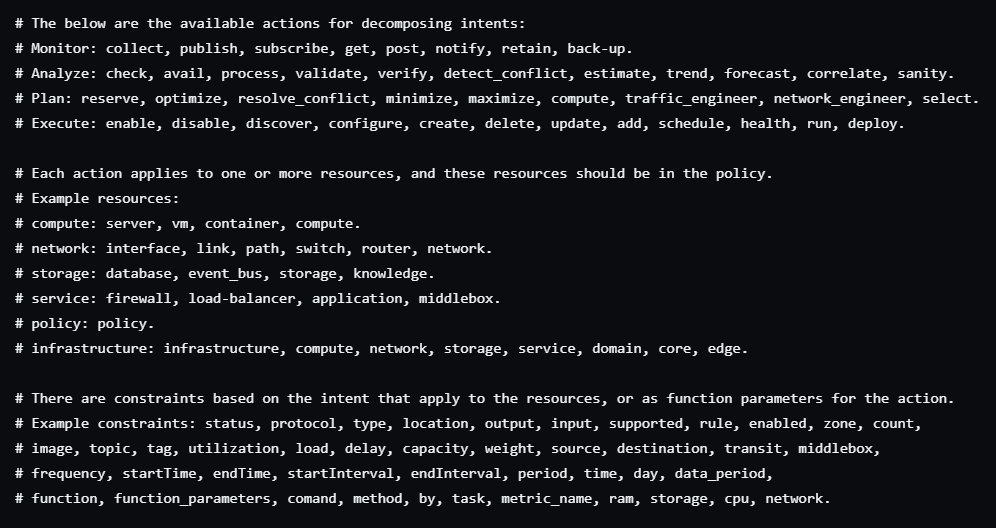} %examples.PNG
   \caption{Example policy actions, resources, and constraints.}
   \label{examples} 
\end{figure}

\begin{figure}[h!]
   \centering
   \includegraphics[width=\columnwidth]{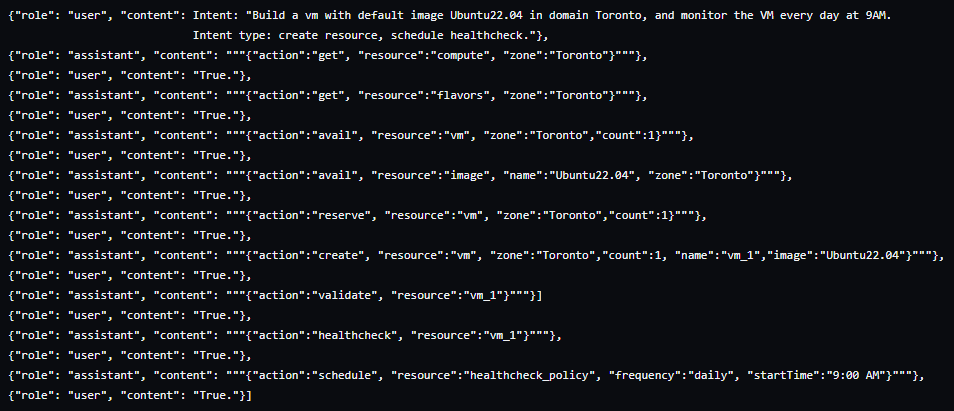} %
   \caption{Few-shot training example for progressive decomposition of an Intent and its corresponding Intent type.}
   \label{training} 
\end{figure}

In the second stage of the pipeline, we test the policy execution in a digital twin environment. We provide the intent, intent type, and generated policies to the validation LLM. Once we have the final version of our policies, if any policies were modified in stage 3, we test the intent again in our digital twin. If there are no issues, we can safely deploy the intent in the production environment. we have a dedicated project for the digital twin within our testbed, so we can test policy deployments without affecting other domains. 

\begin{figure*}[h!]
   \centering
   \includegraphics[width=\textwidth]{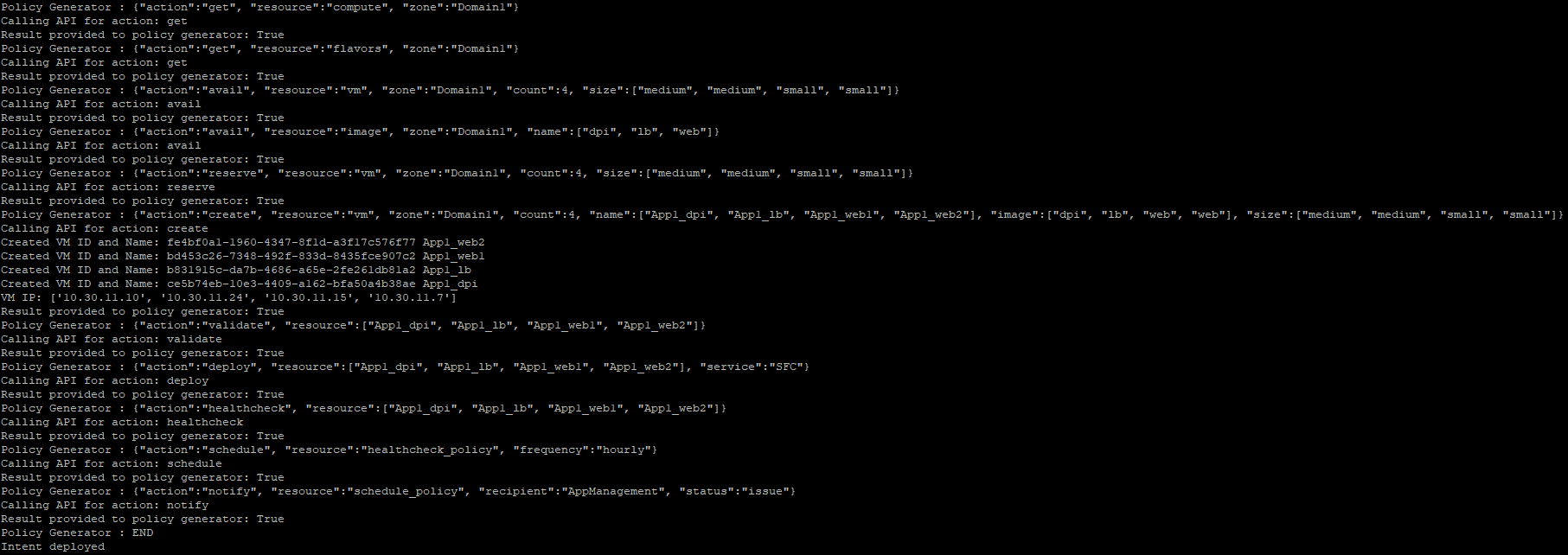} %newresults4-1.png
   \caption{Progressive policy generation and execution for intent fulfillment. Intent: "Deploy a service function chain with high availability in Domain1 consisting of: a medium vm for the dpi service, a medium vm for the load-balancer service, and 2 small vms for the web servers". Intent type: create resource, deploy service, availability.}
   \label{decomposition} 
\end{figure*}

\begin{figure*}[h!]
   \centering
   \includegraphics[width=\textwidth]{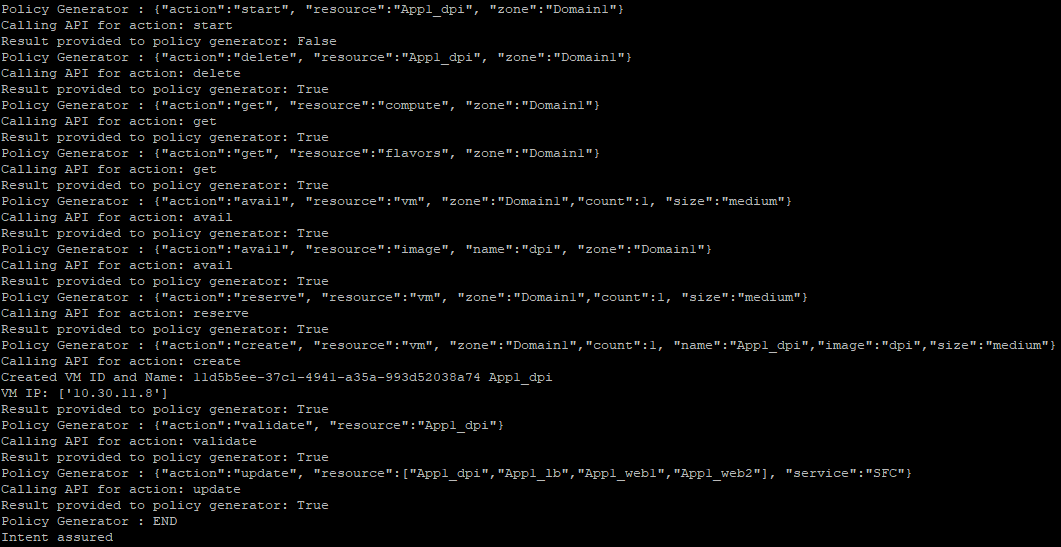} %newresults4-4.png
   \caption{Progressive policy generation and execution for intent assurance (second scenario).}
   \label{assurance} 
\end{figure*}

\section{Evaluation and Use-case}
We evaluated the pipeline for intent-to-policy decomposition and execution in an intent use-case with intent fulfillment and assurance. The intent in the use-case is to deploy a service function chain that consists of Deep Packet Inspection, load-balancer, and two web servers. The intent is: \textit{"Deploy a service function chain with high availability in Domain1 consisting of: a medium vm for the dpi service, a medium vm for the load-balancer service, and 2 small vms for the web servers."} 

We conduct our experiment using OpenStack in the SAVI testbed which is composed of multiple domains, projects and regions. For example, the region we use to deploy the intent has more than 23 physical servers that provide 800 vCPUs, 4TB RAM, and close to 200TB storage. We performed 5 trials for the fulfillment, and for the assurance, and we report on the average time to fulfill and assure the use-case intent. We use the ChatGPT API (GPT3.5 and GPT4 models) for the LLM few-shot training and intent pipeline. 

\subsection{Intent Fulfillment}
To fulfill the intent, we pass it through the pipeline in Figure \ref{pipeline}. We first classify the intent and then obtain the following intent types: \textit{create resource, deploy service, availability}. Next, the intent and types are provided to the second stage, and progressively decomposed into policies. For each decomposed policy, we map the policy to the corresponding API, and provide the result of the execution to the LLM. The API is based on the policy action, and policy resource and constraints are provided as parameters to the API. Figure \ref{decomposition} shows the result of this stage. In summary, the LLM outputs policies to gather necessary data, check resource availability, make a reservation request, create and validate resources, deploy the service function chain, and creates policies to enable monitoring for high availability (i.e., schedule a health check, and set up notifications). If these actions complete successfully, the intent can be safely deployed (given that the validation stage also completes successfully).

\subsection{Intent Assurance}
To demonstrate assurance, we intentionally shut off the DPI VM to trigger the assurance. This was captured by the health check, which sent a notification about the state of the monitored VMs to the Application Manager module (AppManagement). This module detects the intent drift (the state of the DPI VM is "Shutdown", as opposed to "Running"), and passes this message to the LLM to fix the intent. To test the assurance capability of the LLM, we test two scenarios. In the first scenario, we let the LLM successfully perform a start action for the DPI VM. In this case, two policies are executed: start the VM, and validate the VM. In the second case, we provide a negative result when trying to perform the start action, and in turn the LLM generates additional assurance-related policies: delete the VM, get necessary data and check resource availability again, make a reservation request, create and validate the VM, and last, update the service function chain. The result of the second scenario is shown in Figure \ref{assurance}. 

\begin{table}
\caption{Intent Evaluation Analysis}
\centering
\begin{tabular}{|c|c|c|}
\hline
& Execution Time (s) & Number of policies \\
\hline
Intent Fulfillment & 338.9 & 11 \\
\hline
Intent Assurance 1 & 13.2 & 2 \\
\hline
Intent Assurance 2 & 85.7 & 10 \\
\hline
 \end{tabular}
 \label{table:time}
\end{table}

\subsection{Evaluation Results and Analysis}
The results from fulfillment and assurance are shown in Figures \ref{fulfillmentresult} and \ref{assuranceresult} respectively. 
We report the average times and the number of generated policies to fulfill and assure the intent (for both assurance scenarios) in Table \ref{table:time}. The execution results mostly depend on the time our testbed APIs take to complete (e.g., create VM). The time to receive a response using the ChatGPT API is mostly negligible. Although not immediately evident, the policies help us quantify the hidden complexities when working with intents. For example, to fulfill the use-case intent, 11 policies were generated, and these policies are essentially defining the main logic for how to realize the intent. Considering that the policies get mapped to API calls, this means our policies abstract even more lines of code executed through the API calls. 

One important aspect of an intent-based system is the ability to generalize well, meaning that the system can handle modified and unseen intents. From our evaluations, we find that the LLM can handle this requirement well. For example, the intent used for the few-shot example was simpler and had different requirements compared to the use-case intent. However, the LLM was able to learn from the provided context (few-shot learning), such that it can generalize and decompose intents with unseen requirements. These results indicate that generic LLMs are very promising for intent-based applications and management.  

\begin{figure}[h!]
   \centering
   \includegraphics[width=\columnwidth]{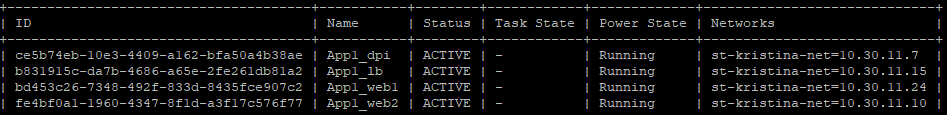} %newresults4-2.png
   \caption{Result of the intent fulfillment.}
   \label{fulfillmentresult} 
\end{figure}

\begin{figure}[h!]
   \centering
   \includegraphics[width=\columnwidth]{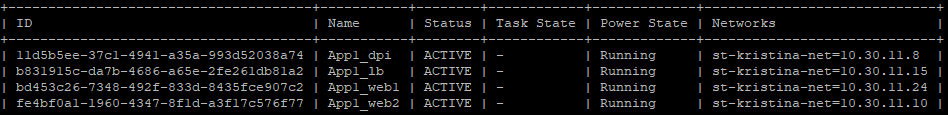} 
   \caption{Result of the intent assurance.}
   \label{assuranceresult} 
\end{figure}

\section{Discussion and Opportunities}\label{few-shot}
We now discuss our insights and additional LLM opportunities for intent-based applications, lessons learned, limitations, and future work. 

\subsection{LLMs for additional conversions}
In this paper, we demonstrated that LLMs can be used to decompose intents into a high-level abstraction, such as a policy. Per our policy framework, the policies specify the actions, resources, and constraints to be applied. However, to execute these policies, we either need to map them to a language that supports implementation, such as YANG (e.g., to the MEF PDO policy model), or to a corresponding API. In this work, we use separate mapping functions in our implementation to map policies into APIs. However, an LLM can also be trained to be able to convert policies into other formal languages, including programmable code, e.g., Python APIs. The benefit of deriving the policies first, is that it allows us to obtain the intent workflow, i.e., the set of steps to fulfill or assure intents in a MAPE loop. 

In our prior work (Emergence), we use a set of pre-defined Finite State Machines (FSM) to execute policies. However, it would be more beneficial to be able to dynamically generate these FSMs. In this manner, the system would be able to execute new, unseen intents, provided that the required functions (e.g., APIs) are available. One approach to creating dynamic, on-demand FSMs, would be to train a generic LLM to create Boolean-Logic Decision Trees from the policies, and then convert these trees into Finite State Machines.

\subsection{Lessons Learned}
An important consideration for LLMs with few-shot learning include over-training and under-training. It is worth noting that we do not modify the weights of the model, but instead provide descriptions, definitions, and usage examples as input to the model, so that the underlying Transformer architecture can learn to decompose intents into policies. The goal is to be able to train the model to generalize well, in order to consistently generate policies that are correct and work as intended. To help generalize better, before a new intent is received, we clear the conversation history, and we send the training prompt to the model. If the training prompt contains too many details, it can confuse the LLM, or it can cause the LLM to output exactly as per the prompt (even if incorrect). If there is too little information, the LLM can begin to deviate and create new attributes or entities (e.g., actions, constraints, resources) that are irrelevant to the system. 

In this paper, we provide simple results back to the LLM upon the execution of a policy, for example, a Boolean value (True, False). If more information is provided to the model, such as some details from the execution, then the LLM could modify the policy constraints accordingly in an attempt to meet the intent. For example, if a policy asks for a large size VM, or a specific version of an image, and these are not available, then, if the LLM receives information about other available sizes and images as part of the result, the LLM will select the best option, e.g., a medium VM, and an image version as close as possible to the original intent version requested. This shows that we can further improve the reasoning logic for the LLM, for example by embedding specific algorithms as part of the few-shot training. In this way, the LLM can play an even bigger role in management.

\subsection{Limitations and Future Work}

Although we demonstrated that a generic large language model can be very efficient in decomposing intents using few-shot learning, there are a number of challenges to consider:
\begin{itemize}
    \item Validation is important to ensure that the policies capture the desired behaviour. This requires validating the policies in terms of the format, proper sequence, correctness and omissions of attributes. For this, we can train an LLM to look for these types of issues. Second, we need to check for logic and algorithmic issues. For this, we can leverage FSMs to ensure correct state transitions, and that the final state matches the intent's desired state. Finally, it is desirable to use a combination of simulators and emulators to create a real digital twin environment.
    \item The transformer model comes with context length limitation, i.e., the number of tokens that the model can attend to. This limits how far the transformer can look back in the conversation history. As such, this also limits the size of the training prompt. However, recent models are increasing the context limit, e.g., GPT3.5 supports 4096 tokens, and GPT4 variants offer 8k or 32k.
    \item The Boolean results provided to the LLM can inadvertently cause the LLM to relax some important constraints, or even start generating policies that deviate from the user's intent (i.e., beyond the scope) as the LLM tries to "desperately" meet the intent. Therefore, it is important to carefully craft the intent, for example by using specific pointer words such as \textit{shall, must, should} to "limit" the LLM's efforts. For example, consider an intent to use a path between two nodes with some specific requirements. If such a path is not available currently, the LLM might try to use actions such as engineering the traffic, or even the network, if the previous attempt fails. This is a major overreaction to a simple intent.
\end{itemize}

\section{Conclusions}
In this paper, we focused on leveraging a generic LLM in our Emergence system for intent-based management and applications. Specifically, we proposed using LLMs with few-shot learning to enable progressive policy generation, driven by policy execution results. We considered both intent fulfillment and assurance, and our results indicate that LLMs are very promising in attaining our goal to enable automatic intent decomposition for application management.

%\nocite{*}
\bibliographystyle{IEEEtran}
\bibliography{IEEEabrv,references}

\end{document}